\documentclass[11pt]{article}
\usepackage{epsfig}
\begin{document}
\title{Effect of Surface Elasticity on Mode-3 Crack Stress Fields}
\author{Tarun Grover$^a$}
\maketitle
\noindent {$^a$  Mechanical   Engineering  Department,  I.I.T.-
Kanpur, U.P.  - 208016,  \newline INDIA }

\begin{abstract}
\noindent  The surface stress can have important effects on the elastic properties of nano-sized structures. Here we analyze the effect of surface stress on Mode-3 crack displacement and stress-field solutions under the assumption of linear elasticity. We show that surface effects generate non-$K$ terms near the crack-tip. We also find the effect of such terms on the conventional small-scale yielding assumption.   
\end{abstract}
\section{Introduction}
The distinction between the energy involved in creating a new surface, as compared to stretching an already existing one, has been reemphasized many times over the years \cite{gibb06,shut50,mull63,camm94} . While the former can fairly be taken as a constant per unit area, the latter may have complicated dependence on surface strain. Here we try to find the effect of the latter (`surface elasticity') on crack field solutions when this dependence is quadratic akin to its bulk counterpart.

Physically, this effect will be important when the solid dimensions are of the order of $\kappa /\mu$ where $\kappa$ and $\mu$ are surface and bulk elasticity moduli respectively. For real materials this ratio is of the order of nanometers. Since the dimensions of cracks in many materials such as glass range from few angstroms to micrometers, and crack elastic field solutions depend largely on surface tractions, it seems worthwhile to analyze the effect of surface elasticity on crack field solutions. 
At the outset, the use of continuum elasticity at such dimensions may be questioned. Some years ago Shenoy et al compared the continuum elasticity results with atomistic simulations \cite{mill00} and were able to show that  bending and torsional behavior of a solid at such dimensions can be accounted for by taking surface elasticity into account. 
Here in similar spirit we try to formulate the problem of a mode-3 crack for a material having elastic surfaces.

\section{Elastic solution for a Mode-3 Crack with Surface Elasticity}
In this section we solve for the elastic field of a mode-3 crack for a solid having an elastic surface. We follow a perturbative approach with the field in the absence of surface elasticity as the unperturbed solution. The domain $\Omega$ consists of $0< r <R , -\pi <\theta < \pi$ (see Fig 1). Let the loading at the external boundary $r = R$ be of the form $\tau_{rz}(R,\theta)= \tau_{a} F(\theta)$.

\begin{figure}
	\centering
		\includegraphics{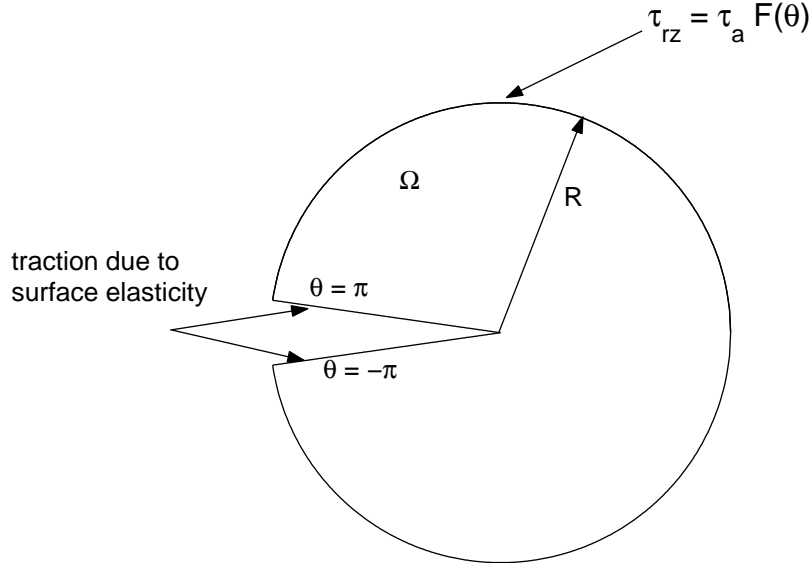}
\caption{\label{fig:domain1} The specimen }
\end{figure}

In the absence of surface tension and surface elasticity effects, the only non-zero displacement $w_o(r,\theta)$ in the z-direction satisfies 

\begin{eqnarray}
  \nabla^2 w_o(r,\theta) & = & 0    \hspace{1cm} (0 < r < R) \label{eq:bulk}  \\
   \frac{1}{r}\frac{\partial w_o}{\partial \theta} & = & 0  \hspace{1cm} (r, \theta = \pi,
 \, -\pi) \label{eq:bc1woel} \\
   \mu \frac{\partial w_o}{\partial r} & = & \tau_{a}F(\theta) \hspace{1cm} (r = R) \label{eq:bc2woel} 
\end{eqnarray}

Supplementing the above equations with the condition of finite displacement at $r=0$, the anti-symmetric solution (i.e. the one which contributes to crack-opening) to above set of eqns. can be easily found as

\begin{eqnarray}
w_{o}(r,\theta) & = & \sum_{n=0}^{\infty} w_{on} \label{eq:w_o1} \\
w_{on} & = & A_{n} r^{n+ 1/2}sin((n+1/2)\theta)  \label{eq:w_on1} \\ 
A_{n} & = & \frac{\tau_{a}F_n}{\mu (n+1/2)R^{n-1/2}} \label{eq:A_n}
\end{eqnarray}

where $ F_{n} = \frac{1}{\pi} \int_{-\pi}^{\pi} F(\theta) sin((n+1/2)\theta) \, d\theta $.

A general formulation of elastic surfaces was given by Gurtin and Murdoch \cite{gurt75}. Following \cite{gurt75}, in the presence of surface elasticity/tension effects, the equation of equilibrium in the bulk remain unchanged (i.e. same as Eq.~(\ref{eq:bulk})) while the boundary conditions at a free surface are changed from $\sigma_{ij} n_j = 0$ to:

\begin{eqnarray}
S_{i\alpha,\alpha} & = & \sigma_{ij} n_j \nonumber \\
S_{\alpha \beta} & = & \sigma_o \delta_{\alpha,\beta} + (\lambda + \sigma_{o}) u_{\gamma,\gamma} \delta_{\alpha,\beta} + (\mu_{o}-\sigma_{o})(u_{\alpha,\beta} + u_{\beta,\alpha}) + \sigma_{o} u_{\alpha,\beta} \nonumber
\end{eqnarray}

Here $S_{\alpha \beta}$ denote surface stress tensor while $\lambda$, $\mu_{o}$ and $\sigma_{o}$ are surface elastic moduli. If the normal to the surface lies along direction 2, then Latin subscripts $i,j,k$ range over 1,2,3 while Greek subscripts $\alpha,\beta$ range over only the directions lying along the surface viz. 1 and 3. The derivatives such as $S_{i\alpha,\alpha}$ or $u_{\alpha,\beta}$ employed in the above eqns.  are surface derivatives i.e. projection of usual derivatives along the surface. Note that $\mathbf{S}$ is linear in surface strain $\mathbf{E^s}$ where $E^s_{\alpha\beta} =  \frac{1}{2}(u_{\alpha,\beta} + u_{\beta,\alpha}) $.
\newline
We switch to cartesian coordinates for a while for sake of mathematical convenience. Along the crack faces ($\theta = \pi,  -\pi$), the negative $x$ axis coincides with $\theta=\pi (-\pi)$ for $y=0^+ (0^-)$.

Substituting $u_{\alpha} = w(x,y) \delta_{\alpha3}$ in above equations

\begin{eqnarray}
\mathit{\mathbf{E^s}} & = &  \frac{1}{2}\frac{\partial w}{\partial x} \mathbf{(e_1 \otimes e_3} + \mathbf{e_3 \otimes e_1)} \label{eq:surfstrain} \\
\mathbf{S} & = & \sigma_{o}\mathbf{e_1 \otimes e_1}  + (\mu_{o} -\sigma_{o}) \frac{\partial w}{\partial x} \mathbf{(e_1 \otimes e_3} + \mathbf{e_3 \otimes e_1)} + {\sigma_o} \frac{\partial w}{\partial x} \mathbf{e_1 \otimes e_3} \hspace{0.5cm}\label{eq:surfstress}
\end{eqnarray}

Taking surface divergence of $\mathbf{S}$ one obtains 
\begin{eqnarray}
 \sigma_{32} n_2 = (\sigma_o \delta(x) + \kappa \frac{\partial^2 w}{\partial x^2})
\end{eqnarray}
where $\kappa =\mu_{o} -\sigma_{o}$.
The first term arises because the constant part of surface stress in Eq.~(\ref{eq:surfstress}) drops abruptly to zero at the crack tip. Its effect on stress fields was solved for many years ago by Thomson et al \cite{thom86} who showed that such terms lead to $1/r$ singularity in stresses at the crack-tip. As we are concerned with the effect of surface elasticity and not surface tension in this work and since the overall field can be found as superposition of the fields due to both terms individually, we drop the first term in above eqn. Keeping this in mind, the above eqn. in $r-\theta$ coordinates may be written as
\begin{eqnarray}
  \frac{1}{r}\frac{\partial w}{\partial \theta} = -\frac{\kappa}{\mu} \frac{\partial^2 w}{\partial r^2} \hspace{1cm} \theta = \pi \label{eq:bc1wel} \\
 \frac{1}{r}\frac{\partial w}{\partial \theta} = \frac{\kappa}{\mu} \frac{\partial^2 w}{\partial r^2} \hspace{1cm} \theta = -\pi \label{eq:bc2wel}
\end{eqnarray}

Eqn. ~(\ref{eq:bulk}) in conjunction with ~(\ref{eq:bc1wel}) and ~(\ref{eq:bc2wel}) completes the specification of boundary value problem for our problem.
We were unable to find a straightforward method to solve this set of equations. 
We will solve for $w$ as a series expansion in terms of the parameter $\frac{\kappa}{\mu R}$. We make the assumption $\frac{\kappa}{\mu R} \ll 1$.

Next write $w(r,\theta)$ as

\begin{eqnarray}
w(r,\theta) & = & \sum_{n=0}^{\infty} w_{n} \label{eq:w_tot1} \\
w_n & = & w_{on}+ (\kappa/\mu) g_{1n} +  (\kappa/\mu)^2 g_{2n} + ... + (\kappa/\mu)^k g_{kn} + ...
\end{eqnarray}
where $w_{on}$ is given by ~(\ref{eq:w_on1}). Throughout our analysis we assume that $\frac{\kappa}{\mu R} \ll 1$. \newline
 Since $w_n$'s are independent, each must satisfy the equation $\nabla^2 w_n = 0$ in the bulk and the boundary conditions ~(\ref{eq:bc1wel}), ~(\ref{eq:bc2wel}) individually. Thus one gets the following set of equations

\begin{eqnarray*}
\nabla^2 g_{kn}(r,\theta) & = & 0    \hspace{1cm} (0 < r < R) 
\end{eqnarray*}

for all $k$, and \newline
\begin{eqnarray}
\frac{1}{r}\frac{\partial w_{on}}{\partial \theta} & = & 0 \nonumber \\ 
\frac{1}{r}\frac{\partial g_{1n}}{\partial \theta} & = & \mp\frac{\partial^2 w_{on}}{\partial r^2} \label{eq:g_1n} \\
\frac{1}{r}\frac{\partial g_{2n}}{\partial \theta} & = & \mp\frac{\partial^2 g_{1n}}{\partial r^2} \label{eq:g_2n} \hspace{1 cm} \lbrace\theta = \pi, -\pi\rbrace  \\
...\nonumber \\
\frac{1}{r}\frac{\partial g_{k}}{\partial \theta} & = & \mp\frac{\partial^2 g_{k-1 n}}{\partial r^2} \nonumber
\end{eqnarray}

where upper sign stands for $\theta = \pi$ and lower for $\theta = -\pi$. Substituting for $w_{on}$ from ~(\ref{eq:w_on1}) in ~(\ref{eq:g_1n}) and solving for $g_{1n}$
\begin{eqnarray}
g_{1n}
& = & (n+1/2)A_{n}r^{n-1/2} cos((n-1/2)\theta) + \alpha_{n}r^{n-1/2}sin((n-1/2)\theta) \nonumber
\end{eqnarray}
where $\alpha_{n}$ is an arbitrary coefficient.
Since the loading due to surface elasticity is symmetric (note the sign change in boundary condition at $\theta=\pi$ and $-\pi$) , all the terms containing $\kappa$ should be symmetric in $\theta$. Therefore $\alpha_{n}= 0$. Thus the terms containing $\kappa$ do not contribute to the crack opening displacement but the stress field is necessarily altered due to surface elasticity. \newline
 Substituting the expression for $g_{1n}$ in ~(\ref{eq:g_2n})
\begin{eqnarray*}
 \frac{1}{r}\frac{\partial g_{2n}}{\partial \theta} & = & 0
\end{eqnarray*}
 Again by the same argument $g_{2n}$ is seen to be identically equal to zero.
It immediately follows that $g_{2n} = g_{3n} = ...= 0$. Thus the perturbative solution is 
\begin{eqnarray}
    w_{n} & = & A_{n} r^{n+ 1/2}sin((n+1/2)\theta)+ (\kappa/\mu)(n+1/2)A_{n}r^{n-1/2}     cos((n-1/2)\theta) \hspace{0.8cm} \label{eq:pert1}
\end{eqnarray}

\section{Small Scale Yielding and Surface Elasticity}
From equations ~(\ref{eq:w_tot1}),~(\ref{eq:pert1}) one observes that the displacement diverges if we include the term $n=0$. As 
 and Ruina \cite{hui95} have shown through their analysis that one needs to be careful while discarding non-$K$ singular(and non-singular) terms for crack problems. In fact, such terms are always present if one take a more realistic situation where there is a nonlinear zone $\Omega'$ surrounding the crack as shown in figure 2.  Here we do an analysis similar to \cite{hui95} to determine the condition for small-scale yielding in the presence of surface effects.   

\begin{figure}
	\centering
		\includegraphics{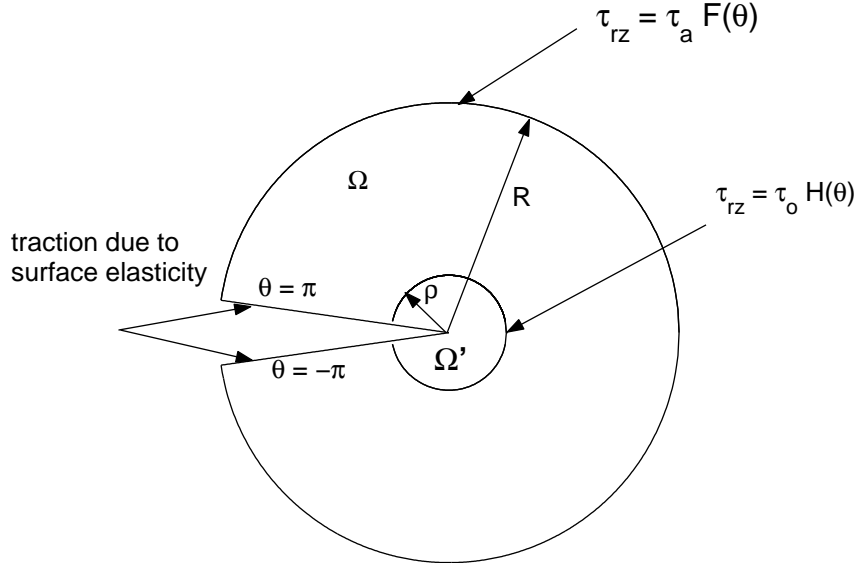}
\caption{\label{fig:domain2} The specimen }
\end{figure}

Assume that the nonelastic zone is centered at the origin $x,y=0$ and has dimensions smaller than the crack length $R$. We now consider a completely elastic problem in the domain $\rho < r <R$, $\rho$ being the outermost radius of $\Omega'$. The traction at the inner radius $r=\rho$ is taken as
\begin{eqnarray*}
\tau_{rz}(\rho,\theta) & = & \tau_{0}H_{\theta}
\end{eqnarray*}
together with the loading $\tau_{a}F(\theta)$ at $r=R$. The antisymmetric solution to the elastic problem in the absence of surface effects can now be found as
\begin{eqnarray*}
w_{o}(r,\theta) & = & \sum_{n=0}^{\infty} w_{on} \\
w_{on} & = & \{C_{n} r^{n+ 1/2} + D_{n} r^{-(n+1/2)}\}sin((n+1/2)\theta)  \\ \nonumber
\end{eqnarray*}
where $ F_{n} = \frac{1}{\pi} \int_{-\pi}^{\pi} F(\theta) sin((n+1/2)\theta) \, d\theta $ and $\epsilon = \rho/R $. 

Following the notation of \cite{hui95}, the above field can be expressed as a sum of two terms: $w^{AI}$ and $w^{AO}$ where $AI$ stands for \textit{A}ntisymmetric, \textit{I}nner traction free and $AO$ for \textit{A}ntisymmetric, \textit{O}uter traction free. For sake of completeness we write the expressions for these fields:

\begin{eqnarray*}
w_{o}(r,\theta) & = & \sum_{n=0}^{\infty} \lbrace w^{AI}_{n} + w^{AO}_{n} \rbrace \\
w^{AI}_{n} & = & \frac{\tau_a}{\mu}\frac{b_{n}^{1}}{(n+1/2)}\lbrack\frac{r^{n+1/2}}{R^{n-1/2}} + \frac{\rho^{2n+1}}{R^{n-1/2}r^{n+1/2}}\rbrack sin((n+1/2)\theta) \\
w^{AO}_{n} & = & \frac{\tau_0}{\mu}\frac{b_{n}^{2}}{(n+1/2)}\lbrack\epsilon^{n+3/2}\frac{r^{n+1/2}}{R^{n-1/2}} + \epsilon^{n+3/2}\frac{R^{n+3/2}}{r^{n+1/2}}\rbrack sin((n+1/2)\theta) \nonumber \\
\end{eqnarray*}

with $\epsilon = \rho/R $, \newline
and \newline
$b_{n}^{1} = F_{n}/\lbrack1-\epsilon^{2n+1}\rbrack$ with $ F_{n} = \frac{1}{\pi} \int_{-\pi}^{\pi} F(\theta) sin((n+1/2)\theta)\, d\theta $, \newline
and\newline
$b_{n}^{2} = -H_{n}/\lbrack1-\epsilon^{2n+1}\rbrack$ with $ H_{n} = \frac{1}{\pi} \int_{-\pi}^{\pi} H(\theta) sin((n+1/2)\theta) \,d\theta $, \newline
In the presence of surface effects the boundary conditions ~(\ref{eq:bc1woel}),~(\ref{eq:bc2woel}) remain unchanged except that now our domain is $\rho < r <R, \theta= \pi, -\pi$. Following exactly the same strategy as before, we obtain the following perturbative solution:

\begin{eqnarray*}
w(r,\theta) & = & \sum_{n=0}^{\infty} w_{n} \\
w_n & = & w^{AI}_{n} + w^{AO}_{n}+ \frac{\tau_a}{\mu}(\kappa/\mu)b_{n}^{1}\lbrack\frac{r^{n-1/2}}{R^{n-1/2}}cos((n-1/2)\theta) + \frac{\rho^{2n+1}}{R^{n-1/2}r^{n+3/2}}\rbrack \times  \nonumber \\
&  & cos((n+3/2)\theta)+ \frac{\tau_o}{\mu}(\kappa/\mu)b_{n}^{2}\lbrack\epsilon^{n+3/2}\frac{r^{n-1/2}}{R^{n-1/2}}cos((n-1/2)\theta) + \nonumber \\ & & \epsilon^{n+3/2}\frac{R^{n+3/2}}{r^{n+3/2}}\rbrack cos((n+3/2)\theta) \nonumber \\
\end{eqnarray*}

From this, the stress fields are readily obtained,

\begin{eqnarray*}
\tau_{rz} & = & \mu \frac{\partial w}{\partial r} \nonumber \\
& = & \sum_{n=0}^{\infty}\lbrack \lbrace \tau_{a}b_{n}^{1}\lbrack\frac{r^{n-1/2}}{R^{n-1/2}} - \frac{\rho^{2n+1}}{R^{n-1/2}r^{n+3/2}}\rbrack sin((n+1/2)\theta) \rbrace
+ \lbrace \tau_a(\kappa/\mu)b_{n}^{1} \times \nonumber \\ & & \lbrack(n-1/2)\frac{r^{n-3/2}}{R^{n-1/2}}cos((n-1/2)\theta) - (n+3/2)\frac{\rho^{2n+1}}{R^{n-1/2}r^{n+3/2}}\rbrack cos((n+3/2)\theta) \rbrace \nonumber \\
& + & \lbrace \tau_{o}b_{n}^{2}\lbrack\epsilon^{n+3/2}\frac{r^{n-1/2}}{R^{n-1/2}} - \epsilon^{n+3/2}\frac{R^{n+3/2}}{r^{n+3/2}}\rbrack sin((n+1/2)\theta) \rbrace + \lbrace \tau_{o}(\kappa/\mu)b_{n}^{2} \times \nonumber \\ & &  \lbrack(n-1/2)\epsilon^{n+3/2}\frac{r^{n-3/2}}{R^{n-1/2}}cos((n-1/2)\theta) - (n+3/2)\epsilon^{n+3/2}\frac{R^{n+3/2}}{r^{n+5/2}}\rbrack cos((n+3/2)\theta) \rbrace \rbrack \nonumber \\
\end{eqnarray*}
and \newline

\begin{eqnarray*}
\tau_{\theta z} & = & \mu \frac{1}{r}\frac{\partial w}{\partial \theta} \nonumber\\
& = & \sum_{n=0}^{\infty}\lbrack \lbrace \tau_{a}b_{n}^{1}\lbrack\frac{r^{n-1/2}}{R^{n-1/2}} + \frac{\rho^{2n+1}}{R^{n-1/2}r^{n+3/2}}\rbrack cos((n+1/2)\theta) \rbrace - \lbrace \tau_a(\kappa/\mu)b_{n}^{1} \times \nonumber \\ & &  \lbrack(n-1/2)\frac{r^{n-3/2}}{R^{n-1/2}}sin((n-1/2)\theta) + (n+3/2)\frac{\rho^{2n+1}}{R^{n-1/2}r^{n+3/2}}\rbrack sin((n+3/2)\theta) \rbrace \nonumber \\
& + & \lbrace \tau_{o}b_{n}^{2}\lbrack\epsilon^{n+3/2}\frac{r^{n-1/2}}{R^{n-1/2}} + \epsilon^{n+3/2}\frac{R^{n+3/2}}{r^{n+3/2}}\rbrack cos((n+1/2)\theta) \rbrace
- \lbrace \tau_{o}(\kappa/\mu)b_{n}^{2} \times \nonumber \\ & & \lbrack(n-1/2)\epsilon^{n+3/2}\frac{r^{n-3/2}}{R^{n-1/2}}sin((n-1/2)\theta) + (n+3/2)\epsilon^{n+3/2}\frac{R^{n+3/2}}{r^{n+5/2}}\rbrack sin((n+3/2)\theta) \rbrace \rbrack \nonumber \\
\end{eqnarray*}

From above equation the condition for $K$ ($\frac{1}{\sqrt{r}}$ term) field dominance can be determined. Comparing the total stress field due to $K$ field  to that of next singular $\frac{1}{r \sqrt{r}}$ term one obtains the following expression for the region in which $K$ field dominates:

\begin{eqnarray}
r & > & \frac{\lbrack(\tau_{a}/\tau_{o})(\rho + a_{o}) + \epsilon^{3/2}(R+a_{o})\rbrack}{\lbrack(\tau_{a}/\tau_{o})(1 + a_{o}/R) + \epsilon^{3/2}(1 + a_{o}\epsilon/R)\rbrack} \label{eq:region} 
\end{eqnarray}               

where $a_{o}= \kappa/\mu$ \newline                                                                
It immediately follows from above eqn. that when $\tau_{o} = 0$, the K-dominance region diminishes in its radius by $a_{o}$ while for the case when $\tau_a = 0$ the $\frac{1}{r \sqrt{r}}$ term always dominates the $K$ term as it does in the absence of surface elasticity.

\section{Discussion}

To put our analysis in perspective, many articles pertaining to effect of surface stress on crack field solutions have appeared in the past \cite{raja73, thom86, wu99}. The main difference with our work is that we have considered surface stress $S$ as having linear (and thus surface energy quadratic) dependence on surface strain $E^s$ instead of being independent of it. As mentioned earlier that a constant surface stress leads to $r^{-1}$ singularity for a Mode-3 crack \cite{thom86}. We have shown that the linear term causes a $r^{-3/2}$ singularity at the crack-tip for $\epsilon=0$.
Physically speaking, non-$K$ singular terms due to surface elasticity should be dominant only in the region of extent $a_o = \kappa/\mu$. Indeed we see from Eq. ~(\ref{eq:region}) that surface elasticity decreases the extent of $K$-dominant region by $a_o$.




Though we are not aware of any direct experimental evidence which favors a quadratic dependence of $U_s$ on $E^s$ over a linear one, Miller et al \cite{mill00} showed that a quadratic dependence was consistent with atomistic simulations for simple problems of elongation, bending and torsion. Note that the variable $S$, which is responsible for size-dependent elastic properties in \cite{mill00}, has same significance as $\kappa$ in this article under Gurtin-Murdoch's formalism \cite{gurt75}.
Finally, the Green's function obtained by Thomson et al \cite{thom86} for symmetric dipole acting along a Mode-3 crack's surface may also be used for obtaining the results in this article but since in our problem the loading itself is dependent on the displacement field, we found it more expedient to directly solve the equations of elasticity instead of a solving a singular integro-differential equation.\newline

\noindent\textbf{\underline{Acknowledgment}}: I would like to thank Dr S Basu, Department of Mechanical Engineering, for stimulating discussions and helpful suggestions.

\end{document}